\documentclass{article}
\usepackage[utf8]{inputenc}
\usepackage{graphicx}
\usepackage{bm}
\usepackage{textcomp}
\usepackage[font=bf]{caption}
\usepackage{amsmath}
\usepackage{float}
\usepackage[margin=1.2in]{geometry}
\usepackage[T1]{fontenc}
\usepackage{kpfonts,baskervald}
\usepackage{lipsum}
\usepackage[labelfont=bf]{caption}
\usepackage{hyperref}
\usepackage{setspace}
\usepackage[style=apa, 
backend=biber, url = false, doi = true]{biblatex}
\usepackage{algorithm}
\usepackage{algorithmic}
\usepackage{authblk}
\usepackage[]{mdframed}

\hypersetup{
    colorlinks=false
}   
\AtEveryBibitem{\clearfield{month}}
\AtEveryBibitem{\clearfield{day}}

\author[1]{Brandon Caie}
\author[1]{Gunnar Blohm}
\affil[1]{Centre for Neuroscience Studies, Queen's University}
\begin{document}
\title{History-dependence shapes causal inference of brain-behaviour relationships}
\date{}
\maketitle
\section*{Abstract}

Behavioural and neural time series are often correlated with the past. This history-dependence may represent a fundamental property of the measured variables, or may arise from how confounding variables change over time. Here we argue that undecidability about the ground-truth of history-dependence is a general computational property of systems that exchange information with its environment, and show that the resulting uncertainty has a direct impact on causal inference. We first argue that uncertainty in the ground truth of history-dependence is an inherent property of open systems that cannot be explicitly falsified. Simple model systems are then simulated to show how different assumptions about history-dependence can lead to spurious correlations and statistical properties of data distributions that are typically  unaccounted for. We then consider this problem from an interventionist perspective, showing that interventions can only be guaranteed to remedy the spurious correlation problem when the latent dynamics between the intervention and measured processes are known \textit{a priori}, and the effect of the intervention is invariant at the chosen level of analysis. We conclude that uncertainty about history-dependence is a fundamental property of the study of neural systems, and in light of this discuss how causality should be assessed in neuroscience.
\newpage
\section*{Introduction}
Behaviour and its neural basis can be understood in terms of change (\cite{hebb_organization_1949, skinner_science_1953, bruner_short_2004, niv_reinforcement_2009}). Linking behaviour to neural activity therefore requires the identification of correlational or causal relationships between neural and behavioural processes that are themselves in flux (\cite{gomez-marin_causal_2017, fox_intrinsic_2007, linkenkaer-hansen_long-range_2001, dijkstra_cognitive_2016}). Generally, how does one determine the statistical properties of how a system changes over time? In isolated systems, where there is no exchange of information between the external world, history dependence may be answered by systematically manipulating the inputs and recording outputs -- if one can change the stimulus eliciting a behavioural response, or electrically stimulate a neuron, one can test how the output at one timepoint is dependent on the input from a previous timepoint. \\

When neural and behavioural processes are measured, even with careful controls, there is irreducible variability in observed responses (\cite{carpenter_neural_1999, deco_role_2008, stein_neuronal_2005, faisal_noise_2008, allen_evaluation_1994, renart_variability_2014, waschke_behavior_2021, bell_behavioral_2014, branco_probability_2009}). Because of this, models accounting for behaviour and/or neuronal activity often introduce a random process at some level in order to explain empirical variability (e.g. \cite{deco_stochastic_2009, churchland_variance_2011, kanashiro_attentional_2017}). Once a random event is required to explain uncertainty in any sort of measurement, one may then be interested in where this randomness originates, and what purpose it serves \footnote{In The Foundation of Statistics, \cite{savage_foundations_1954} wrote "It is unanimously agreed that statistics depends somehow on probability. But, as to what probability is and how it is connected with statistics, there has seldom been such complete disagreement and breakdown of communication since the Tower of Babel."} \footnote{Bertrand Russell said that "Probability is the most important concept in modern science, especially as nobody has the slightest notion what it means." \cite{bell_development_1945}}. A random event may be interpreted as an inherent property of the system -- alternatively, a random process may describe variability that arises from a collection of unobserved, but knowable, causes (\cite{kiureghian_aleatory_2009, hullermeier_aleatoric_2021})). In the former case, variability is often considered to be irreducible, and as such a fundamental property of the system, while in the latter case it may arise from dependencies that while theoretically knowable can be sufficiently explained by a collection of random events. \\

One way in which observed variability may be explained through a knowable cause, in lieu of appealing to inherent randomness, is history-dependence (\cite{marcos_neural_2013, kording_dynamics_2007, dhawale_role_2017, lueckmann_can_2018, fischer_serial_2014}). In neuroscience, it is common to collect many trials of a similar behaviour, upon which one will find variability (\cite{gold_neural_2007, huk_beyond_2018}) -- the study of history-dependence asks to what degree previous trials are predictive of future states, and as such a predictor of future variance. Even in cases where there is no apparent benefit to treating trials as independent, it is common to observe statistical dependencies between neural activity and behaviour across-trials (\cite{fecteau_sensory_2004, lau_dynamic_2005, gao_sequential_2009, st._john-saaltink_serial_2016, lange_prestimulus_2013, sugrue_matching_2004, lueckmann_can_2018, macke_choice_2019}). When deconstructing the history-dependence of data that is traditionally aligned to a common event in a repeated task and treated as an ensemble, one is essentially asking to what degree observed variability can be explained by considering a given process not as a collection of independent events, but as different elements of unique sequences. However, in order to test many of these hypotheses about what may be thought of as the essentialism of variability, that is to consider variability as a natural kind, the ability to distinguish stationary randomness from history-dependence needs to be a computationally tractable problem \footnote{Ernst Mayr describes essentialism as a belief system wherein " a limited number of fixed, unchangeable "ideas" underlying the observed variability [in nature], with the eidos (idea) being the only thing that is fixed and real, while the observed variability has no more reality than the shadows of an object on a cave wall" (quoted in \cite{sober_evolution_1980})} . \\

Here, we argue that falsifying the ground-truth history-dependence of neural systems is computationally undecidable. In doing so, we propose a fundamental limitation on the ability to falsify the distinction between inherent randomness and variability arising from history dependence. Importantly, rather than simply providing a limitation on our understanding of a generative model of a living organism, we claim that uncertainty about the ground-truth history-dependence of a system has material consequences on causal inference. By studying simple model systems where history-dependence can be quantified, we arrive at the following primary claims:

\begin{mdframed}[linewidth = .5mm]
    \begin{quote}
        \begin{enumerate}
            \item There is \textit{no free lunch} for the ontological nature of variability in joint observations of neural and behavioural data -- no agnostic vantage from which one can avoid making assumptions about whether a repeated measurement is inherently random or a product of history-dependence.
            \item Inductive biases about variability have direct effects on hypothesis testing and causal inference.
        \end{enumerate}
    \end{quote}
\end{mdframed}  

\subsection*{Undecidability of history-dependence in open systems}

Experimentalists observe statistical dependencies between neural activity and behaviour of previous and current trials (\cite{fecteau_sensory_2004, lau_dynamic_2005, gao_sequential_2009, st._john-saaltink_serial_2016, lange_prestimulus_2013, sugrue_matching_2004, lueckmann_can_2018, macke_choice_2019}). Given this knowledge, one may then be interested in using information derived from statistical dependencies to deduce general properties about the systems that produce observations of behaviour and neural activity. Fundamentally, this is a systems identification approach that seeks to uncover the dynamical properties of the system in question from repeated measurements. However, systems identification requires the ability to control the initial conditions of the system under different controlled inputs. Neural systems, on the other hand, are rarely (if ever) able to be isolated under fixed conditions. Thus, the ability to perform systems identification in neuroscience requires that the dynamics under question are sufficiently stationary and isomorphic across different individuals and/or experimental demarcations (be them time points, trials, sessions, etc) such that fundamental properties can be uncovered. It can be debated if, when and to what degree this may be possible in different biological systems (\cite{marder_variability_2011, molenaar_manifesto_2004, edelman_degeneracy_2001, safaie_preserved_2023}). But how would one go about verifying the stationarity hypothesis in even the simplest of model systems? \\

Consider a simple systems identification problem. Here, an experimenter is trying to determine if a binary sequence generator is governed by a random process, or a deterministic dynamical system (\textit{Fig 1}). The experimenter decides to run 100 experiments, where in each 100 outcomes are recorded. In each of these 100 experiments, a different sequence is observed each time. After averaging all of the outcomes, the probability of observing a 0 is close to that of a 1, so it may be concluded that the underlying process is random. However, for every possible experimental outcome, it would be trivial to construct a deterministic machine, the length equalling the sum of the experiments producing the data, that yields an identical output. Following this, one may repeat another 100 experiments, upon which one could falsify the previous machine, and then propose a new deterministic machine to account for all of the experiments observed thus far, and so on \textit{ad infinitum}. \\

\begin{figure}[!ht]
\centering
\includegraphics{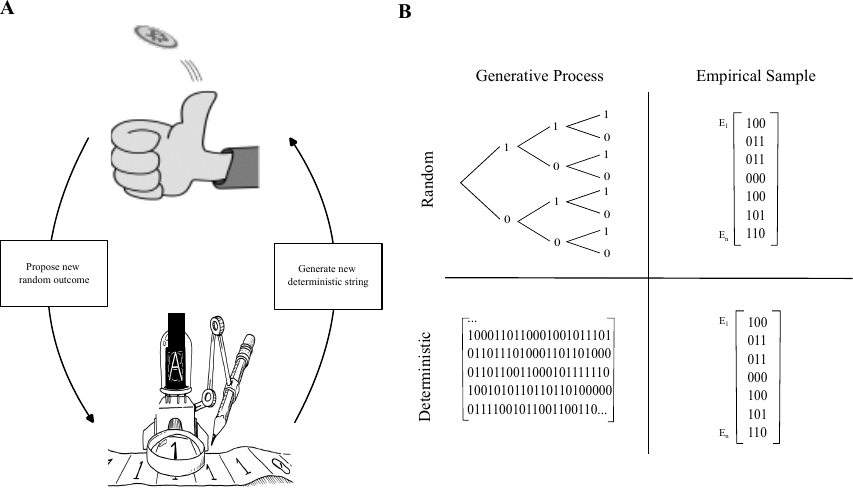}
\captionsetup{justification=raggedright,singlelinecheck=false}
\caption{Undecidability of binary outcomes. \textbf{A: }A program that continues if a deterministic sequence can be constructed to explain the outcome of a random event, upon which a new random event is draw, will never halt \textbf{B:} Two distinct generative process for a binary sequence (random and deterministic) are depicted in the left column, with two identical empirical samples shown in the right column.}
\end{figure}

This recursive logic (\textit{Fig 1}) is the basis for a few famous computability and logical problems, such as the halting problem, the incompleteness theorem, and the diagonal argument (\cite{turing_computable_1937, prokopenko_self-referential_2019, lipton_georg_2013}). Essentially, this class of arguments relies on demonstrating a set of logical statements that are contradictory and result in a recursive chain of plausible justifications. Consider generally that the goal of reasoning about a system is to improve a belief over it, \textit{Q}. The argument posits that while no reason can be \textit{Q} itself (i.e. rejection of circular logic), no reason is sufficiently justified in the absence of a further reason. Applying this logic to deciding whether a system is inherently random or exhibiting history-dependence, explaining variability in data through a random process may always be contradicted by a suitably chosen epistemic source, i.e. a past sequence of inputs that must be retested to falsify, and vice versa. \\

So when assessing history-dependence in biological systems, where precise test-retesting under a fixed history of inputs may not be realizable, one must begin with the premise that establishing the ground-truth history-dependence of a system is computationally undecidable. While a statistical relationship between past and present inputs to the system may be deduced, we should be wary of positioning the recovery of a ground-truth understanding of system dynamics as a goal, even with the ability to sample from it infinitely. Instead, we argue it should be \textit{expected} that an irreducible uncertainty exists in our understanding of any governing dynamics under observation. \\

What are the consequences of this fundamental limitation in the accuracy of systems identification on open systems, where there is a fundamental asymmetry between past and future, such as those likely to explain the neural control of behaviour? Our view is that, beyond providing a barrier to a complete ontology of a living thing, uncertainty over the history-dependence in a system has a tangible and quantifiable effect on statistical inference of causal interventions to neural circuits. Specifically, we will show that statistical testing of brain-behaviour relationships, whether through correlational studies or causal interventions, is dependent on the history-dependence of the system, which is only testable by expanding the timescale over which possible dynamics are considered, thus requiring new experiments to falsify a new set of dynamics that cover the expanded hypothesis space, that in turn require retesting, \textit{ad infinitum}. \\

\section*{Results}

\subsection*{History-dependence in simple model systems}

To build an understanding of the statistical challenges of history-dependent data, we will focus on the study of simple model systems where history-dependence can be quantified and controlled. Maybe the simplest example of a history-dependent system is a random walk. Random walks are a fundamental concept in mathematics and statistics, serving as a useful model for various biological phenomena (\cite{codling_random_2008}). Essentially, a random walk is a mathematical process that describes the trajectory of a particle as it takes successive, random steps in space. In the simplest case, being a one-dimensional random walk, a particle starts at an initial position \(x_{0}\), and at each time step moves left or right by some fixed distance. The direction of the step is determined by a random process, such as a binary outcome in a discrete setting, where 1 represents a step to the right, and 0 a step to the left. The position of the particle at any given time \(t\) is denoted as \(x_{t}\), and can be expressed as the cumulative sum of the random steps taken up to that point

\begin{equation}
x_t = x_0 + \sum_{i=1}^{t} Z_i
\end{equation}

Where:
\begin{align*}
x_t & \text{ is the position at time } t \\
x_0 & \text{ is the initial position at time } 0 \\
Z_i & \text{ are independent and identically distributed random variables representing the steps taken at each time step.}
\end{align*}

Now consider that this process is repeated many times, such that we have a collection of random processes \(X\) composed of these individual walks. \(X_{t}\) can be described as a collection of states at each time point:

\begin{align}
X_t & = (X_t^{(1)}, X_t^{(2)}, \ldots, X_t^{(N)}) \\
& = (X_0^{(1)}, X_0^{(2)}, \ldots, X_0^{(N)}) + \left(\sum_{j=1}^{t} Z_j^{(1)}, \sum_{j=1}^{t} Z_j^{(2)}, \ldots, \sum_{j=1}^{t} Z_j^{(N)}\right)
\end{align}

Here:
\begin{align*}
X_t & \text{ is the state vector representing the entire random process at time } t \\
N & \text{ is the total number of individual random walks in the collection.}
\end{align*}

The expected value of a collection of random walks is always zero, because random walks consist of equally likely steps in both positive and negative directions. This symmetrical nature ensures that, on average, the positive steps cancel out the negative steps, resulting in an expected value of zero. Mathematically, 

\begin{equation}
\mathbb{E}[X_t] = \sum_{i=1}^{t} \mathbb{E}[Z_i] = 0
\end{equation}

However, the variance of the random walk does not stay constant across time. The variance of the random walk is given by:

\begin{equation}
\text{Var}[X_t] = \sum_{i=1}^{t} \text{Var}[Z_i] = t \cdot \sigma^2
\end{equation}

Where \(\text{Var}[X_t]\) represents the variance of the random walk at time \(t\), and \(\sigma^2\) is the variance of each individual step. Because the variance of the ensemble  \(X\) grows with time, while the expected value remains zero, we can deduce something important about its properties: a random walk is non-ergodic. An ergodic process is one wherein the statistical properties derived from one long realization are the same as those obtained from an ensemble of many independent realizations. In short, an ergodic process allows you to learn about the ensemble behavior by observing a single, infinitely long sample path, and vice versa. From an experimentalist's perspective, knowing a given measurement comes from an ergodic process can be important, because it allows for knowledge about many measurements of a system collected at one moment in time to generalize to how single measurements change with time, and vice versa.  \\

A random walk is non-ergodic because it does not satisfy this property. In a random walk, the statistical properties obtained from a single trajectory can differ from those derived from an ensemble of independent realizations. This discrepancy arises due to the inherent randomness in individual steps and the accumulation of that randomness over time. As a result, any given random walk may exhibit a wide range of possible trajectories, while the behavior observed over a finite time span may not necessarily reflect the entire ensemble (\textit{Fig 2a}). Thus, while the expected mean of the process is unchanging across time, individual trajectories vary considerably, such that the expected variance of the ensemble continues to increase over time, with more and more processes accumulating sequences of events that are individually unlikely to occur. In the language of ergodic theory, \textit{the time-average does not equal the ensemble average}. This distinction is visualized in \textit{Fig 2A} (ensemble average depicted as the cross-section in purple, and the time average the longitudinal-section in green). \\

\begin{figure}[!ht]
\centering
\includegraphics{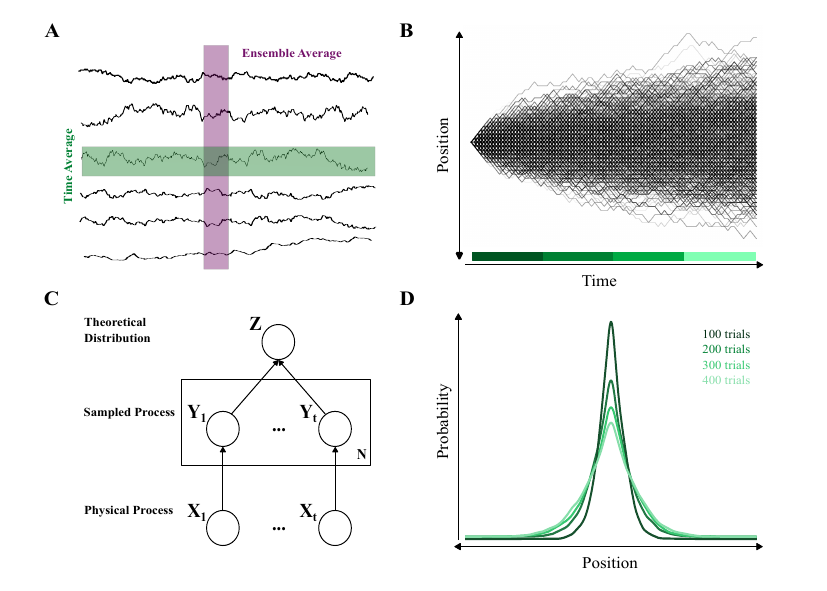}
\captionsetup{justification=raggedright,singlelinecheck=false}
\caption{Non-Ergodicity in a Simple Model of History-Dependence. \textbf{A:}, The ensemble average vs the time average of a set of time series is depicted. In ergodic processes, these properties are equivalent. \textbf{B:}, depiction of a set of random walks with a constant starting position. Non-ergodicity is due to the increase in variance over time with constant mean. \textbf{C:} Plate notation showing a hypothetical problem where an experimenter is sampling from a physical process at different time points. \textbf{D:} Theoretical distribution \textit{Z} for random walk sampling over different lengths of time; note the non-Gaussian heavy tails that increase with experiment length}
\end{figure}

Consider now the study of a system that is governed by a random walk (\textit{Fig 2B}). At every time-point imagine they can sample from a single realization of the process with some variability. For simplicity, assume that the true value of the process is the mean of a normal distribution, where the variance \(\sigma\) of that distribution is the precision with which one may directly sample from the underlying random walk. They may then repeat the experiment N times, such that \(y_{t}\) represents samples from the \(N\) experiments at time \(t\). If the experimentalist could repeat this sampling process over the same random walks an infinite number of times, the theoretical distribution \(Z\) could be recovered by sampling from Y

\begin{equation}
\begin{split}
    Y_t^{(n)} &\leftarrow N(X_t^{(n)},\sigma) \\ 
    Z_t^{(n)} &\leftarrow Y_t^{(n)}
\end{split}
\end{equation}

The theoretical distribution is thus akin to a mixture model, where the probability density of each random walk over Gaussian means represents the theoretical distribution of one process, equally weighted with all other random walks to give the theoretical distribution \(Z\). \textit{Figure 2C} depicts this process in plate notation, where each variable within a 'plate' is repeated 'N' number of times. For this, we simulated the theoretical distribution \textit{Z} for experiments with different numbers of trials (100, 200, 300, 400) in \textit{Fig 2D}.  \\

Two things should be observed in the \(Z\) distributions in \textit{Fig 2D}. First, a greater proportion of the probability density occurs in the tails of the distribution than would be expected from a Gaussian. Secondly, the proportion of the density occupied by the tails is dependent on the length of the experiment. Put simply, this means that the chances of single trajectories showing extreme deviations from the average when accumulating random events increases over time. Thus, while the average across many instances of a random walk  -- the ensemble average-- remains constant, any given realization is not predictive of the ensemble behaviour. Further, the probability of observing a realization that has accumulated far away from the ensemble average increases with the length of an experiment, such that collecting more data from a single trajectory decreases, rather than increases, its predictive power of the ensemble statistics. \\

What are some of the consequences of studying non-ergodic systems? Consider a study that correlates neural activity with a behavioural time series. In absence of knowledge about the true generative model of a system, we can leverage simulations to describe what should be expected in simple systems where we have control over the process. First, consider two independent time series, one a measurement of a behaviour \(B\), and one a measurement of a neural response  \(N\). It has been known since the Monte Carlo simulations of \cite{granger_spurious_1974} (and see \cite{phillips_understanding_1986} for an asymptotic proof) that if these two processes were to follow independent random walks,

\begin{equation}
    \begin{split}
        B_t = B_0 + \sum_{i=1}^{t} Z_i \\ 
        N_t = N_0 + \sum_{i=1}^{t} N_i \\ 
    \end{split}
\end{equation}

while the ensemble of each process have identical means, spurious correlations tend to be found between individual comparisons of the two processes. Informally, this occurs because of the non-ergodicity of random walks -- in a 2D space, one realization of two independent random walks will not cover the space on average equally, so on any given sample a relationship between the two is likely to occur. This has a major consequence for statistical testing that assumes increasing the sample size improves the reliability of rejecting the null hypothesis. In the case of independent random walks, as the number of samples grows, the distribution of correlations remains unchanged, but the distribution under the null hypothesis shrinks -- because of this, collecting more samples from two independent processes increases the probability of observing false positives. \\

Further, the false positive rate between two independent time series can be shown to be dependent on the strength of history-dependence. To show this, we consider an autoregressive (AR) model. An AR model is a stochastic process in which each value in the series is a linear combination of its past values, with random noise added independently at each time step. Mathematically, an AR process of order \(p\) can be expressed as:

\[
X_t = c + \phi_1 X_{t-1} + \phi_2 X_{t-2} + \ldots + \phi_p X_{t-p} + \varepsilon_t
\]

Here:
\begin{align*}
X_t & \text{ represents the current value in the time series. } \\
c  & \text{ is a constant term. } \\
\phi_1, \phi_2, \ldots, \phi_p & \text{ are the autoregressive coefficients } \\
X_{t-1}, X_{t-2}, \ldots, X_{t-p} & \text{ are the lagged values of the time series. } \\
\varepsilon_t & \text{ is a white noise error term with mean zero and constant variance. } \\
p & \text{ is the order of the process} 
\end{align*}

An AR process is therefore a simple model system where the strength of history-dependence can be adjusted parametrically. An AR process is equivalent to a random walk when the autoregressive coefficient \(p\) and the coefficient \(\phi_1\) are equal to 1. In this case, the AR process becomes a first-order autoregressive process, often denoted AR(1), and can be fully described by a random walk. In contrast, an AR(0) process simplifies to a constant term plus white noise, as the series simplifies to a sequence of independent random variables. \\

\begin{figure}[!ht]
\centering
\includegraphics[width = .8\textwidth]{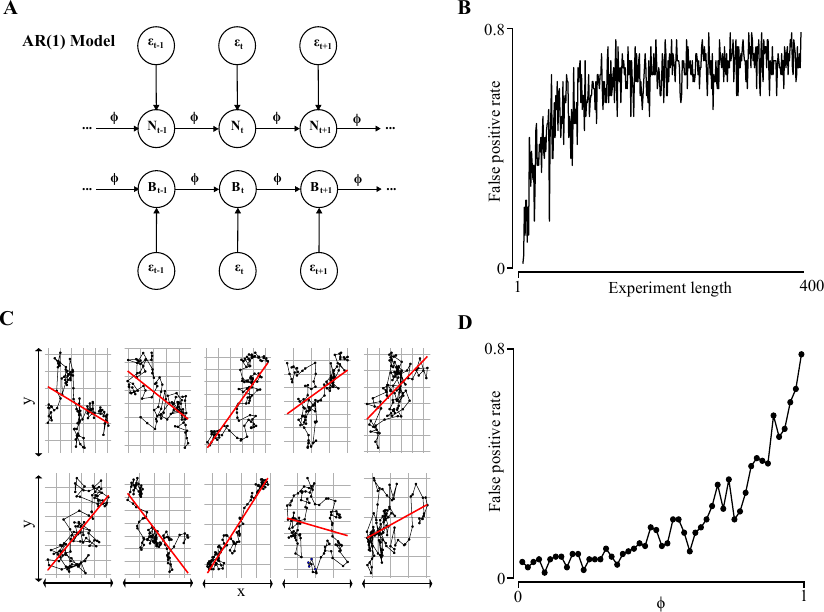}
\captionsetup{justification=raggedright,singlelinecheck=false}
\caption{Spurious correlations in independent processes with history-dependence. \textbf{A:}  Graphical depiction of an autoregressive AR(1) model equivalent to a random walk, where the value of the current timestep is dependent only on the previous value. The strength of the dependency on the previous timestep is controlled by \(\phi\). \textbf{B:} False positive rate for two independent AR(1) processes as a function of the length of the time series. \textbf{C:} Visualization of two independent AR(1) processes as a 2D random walk. 10 random realizations are plotted in black. Regression lines are plotted in red. \textbf{D:} False positive rate for a fixed experiment length of 400 timesteps is plotted against the AR(1) parameter \(\phi\), showing a gradual change in spurious correlations between white noise (\(\phi\) = 0) and a random walk (\(\phi\) = 1).}
\end{figure}

To demonstrate the impact of this history-dependence on false positive rates between two independent time series, we simulated two AR(1) processes while varying \(\phi_1\), and plotted the resulting false positive rate, assuming a significance threshold \(\alpha = .05\). The probability of inferring a spurious relationship between the independent time series was dependent on the strength of the relationship between the previous time-step, as controlled by the \(\phi_1\) coefficient. \textit{Fig 3} summarizes these results. An AR(1) model describing a behavioural and neural process (respectively \(B\), \(N)\)) is depicted in \textit{Fig 3A}, with the false positive rate for different lengths of the time series (here the rate is calculated as the average across 1000 simulations) in \textit{Fig 3B}. In \textit{Fig 3C}, 10 random realizations of the independent \(B\) and \(N\) processes are depicted as a 2D random walk, with the best-fitting linear regression between each shown in red, highlighting the tendency to find a relationship between independent series on any given realization. \textit{Fig 3D} plots the average false positive rate for a fixed time series length (T = 400) across 1000 simulations while varying the \(\phi\) parameter. By systematically varying \(\phi\), we could adjust the AR process to behave like white noise (\(\phi=0\)), a random walk (\(\phi=1\)), and the space in between. In doing so, we can see that the false positive rate for two independent time series is continuously dependent on the strength of history-dependence. Thus, the strength of history-dependence can control the expected correlational structure under a null hypothesis, i.e. two measured processes being independent. 

\subsection*{Mediation analysis over latent intervention dynamics}

Issues surrounding spurious correlations in the data used in neuroscience have been raised previously. Because of this, there have been growing calls for hypothesis testing in neuroscience to be grounded in causal inference (\cite{siddiqi_causal_2022, weichwald_causality_2021, mehler_lure_2018, jonas_could_2017, ross_causation_2024}).  Causal inference methods that seek to discover global causal structures from data are used in neuroscience (\cite{marinescu_quasi-experimental_2018}), such as Granger causality, structural equation models, state-space models, and Bayesian networks. Each of these data-driven approaches have different limitations, and often different understandings of causality, but generally operate by making assumptions about the global structure of an underlying graphical model in order to identify local causal relationships (\cite{pearl_causal_2009}). For the remainder of this work, we will focus on the somewhat distinct interventionist approach to establishing causation, which begins with the problem of identifying local causes in order to uncover an underlying graphical model (\cite{evans_two_2017}). \\

It is first worth mentioning that the logic underpinning causality and experimental interventions in neuroscience is not homogeneous. Some experimental interventions have a great deal of accumulated evidence to act on a certain neural process, and thus become a way to manipulate the neural process to estimate an effect on a behaviour -- in this way, a strong \textit{a priori} understanding of an interventions effect on a biophysical process that we can (indirectly) measure is used to infer the relationship between that biophysical process and an unknown, such as a behaviour. This is what we may call a \textit{neural-deductive} approach, where the path between intervention and neural process is taken as a given in order to deduce a cause between neural process and behaviour. In contrast, one may have an accumulation of evidence that a certain intervention affects a given behaviour, and use measurements of different physical processes to form inferences about how the behaviour is generated -- we may call this a \textit{behavioural-deductive} approach. \\

Unfortunately, a strong \textit{a-priori} understanding of a given intervention is not always realistic. In reality, how a given intervention in cognitive neuroscience is understood is usually based on an accumulation of knowledge across many different levels of analysis. Because of this, the intervention in context of the measured processes often becomes the process of study itself, rather than the intervention existing as a known variable that can be taken as a given (\cite{aalen_causality_2012}). For this reason, and because we believe this issue has been underappreciated in cognitive neuroscience, we will now use model simulations to address the following question: how does adding an intervention to our previous problem of spurious correlations in jointly-measured brain-behavioural data improve our ability to assess causality between jointly measured processes, if there is uncertainty about what the path this intervention takes to a measured variable. Or rephrased, what should our expectations for the null hypothesis be when testing the interaction between new experimental interventions and jointly measured neural and behavioural processes when the underlying graphical model is unknown. And does this depend on the properties of the system(s) it is influencing, i.e. history-dependence, we are measuring? How does uncertainty about these properties change our expectations under the null hypothesis? \\

\begin{figure}[!ht]
\centering
\includegraphics[width=.8\textwidth]{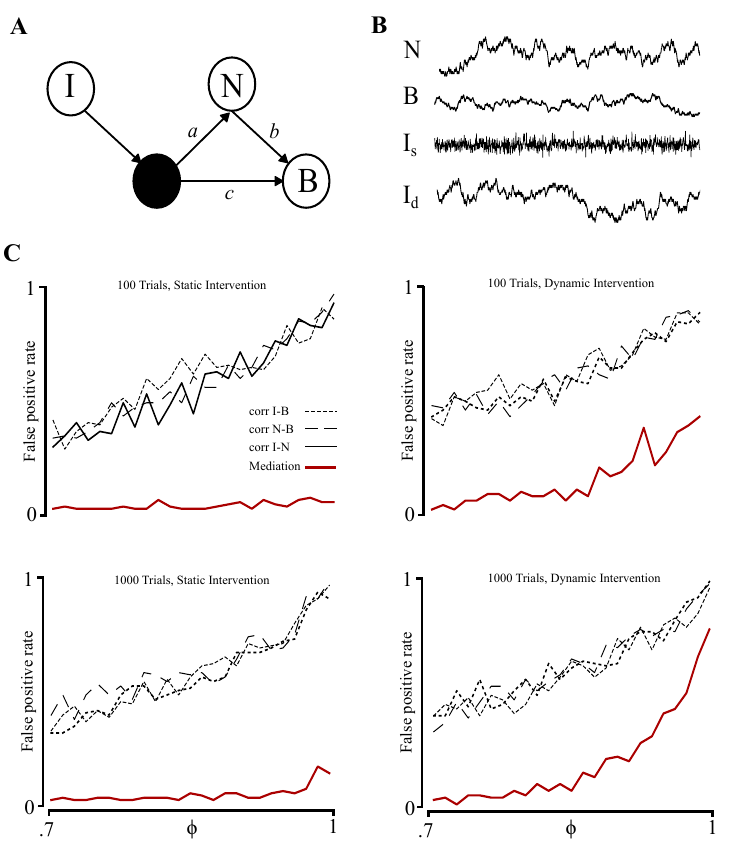}
\captionsetup{justification=raggedright,singlelinecheck=false}
\caption{Latent dynamics between null interventions and measured variables influence mediation analysis. \textbf{A:} Directed graph between an intervention and a neural and behavioural time series through latent dynamics (black circle). Regressions performed for mediation analysis are denoted \textit{a, b, c}. \textbf{B:} Sample time series for \textit{N}, \textit{B}, and \textit{I} under static and auto-correlated latent intervention dynamics. \textbf{C:} False positive rates simulated for 1000 experiments for regression \textit{c} (solid black), \textit{a} (large-dash), and \textit{b} (small-dash), and the mediation sobel test (red). The plots are shown for static (left) and auto-correlated latent dynamics (right) for 100 trials (top) and 1000 trials (bottom) as a function of the \(\phi\) parameter (set to be equal for \(N\) and \(B\)).}
\end{figure}

We now describe a causal intervention model. Typically, an intervention is thought of as an experimentally controlled variable that is known \textit{a priori}. However, when studying complex systems the underlying graphical model between a controlled intervention and a measured process is usually hypothesized, rather than known. While an experiment may have one intervention, and the measurements of one or several neural and behavioural processes, it is not always the case that it is \textit{known} that the intervention directly acts on a measured variable. In the case of high-dimensional systems, such as those dealt with in neuroscience, there are typically many possible paths between an intervention and a causal effect on the measured variable(s). Thus, an experimentally controlled intervention may have a causal effect on a measured process, but often through intermediary processes that may have unknown dynamical properties. \\

For this reason, we ask how statistical analysis of causal interventions may change when different latent dynamics between the intervention as defined by the experimenter and the measured variables are assumed. To do so, we consider the intervention as a dynamical system of the same form as the autoregressive models we have considered previously, and ask what would be expected in the null case where independent neural and behavioural processes are measured over interventions to latent dynamics that do not causally affect each process. We consider a neural \(N(\phi)\) and behavioural \(B(\phi)\) processes that are each generated by an autoregressive model again parameterized by \(\phi\). We then model a causal intervention  over a latent process \(I(\phi)\) that is itself an autoregressive model, where the \(\phi\) parameter can be independently manipulated. It should be re-emphasized that for this null intervention, there is no causal effect between the independently generated \(N\) and \(B\), and \(I\) -- we instead generate each and ask how causal relationships between \(I\), \(N\), and \(B\) may be assessed in the null case where different latent dynamics between an experimentally-defined intervention and the measured processes are assumed. \\

We next require a method to infer causality from the intervention. Mediation analysis is a statistical method to infer a causal path between an independent variable and a dependent variable through the inclusion of a mediating variable (\cite{mackinnon_mediation_2007}). Mediation analysis first confirms the significance of a regression \(c\) between an independent and dependent variable, then performing a regression \(a\) between the independent and mediating variable, and a regression \(b\) between the mediating and dependent variable (\textit{Fig 4A}). Significance testing for mediation analyses can then be performed in several ways, for example a Sobel test, where a mediation effect is significant if both \(c\) supports a significant relationship and a mediation score is significant, using a z-score computed as 

\begin{equation}
    z = \frac{a*b}{\sqrt{b^{2}sa *a^{2}sb}}
\end{equation}

We use this approach to deducing causal relationships to ask if, and under what conditions, knowledge about the latent dynamics between an experimentally controlled intervention and a measured process can reduce the impact of spurious correlations in autocorrelated data. To this end, we simulate \(N\) and \(B\) independently for 1000 experiments, and compute the correlation as a function of \(\phi\), as before, to get a baseline regression for each experiment. However, here we then consider \(N\) as a mediator between \(I\) and \(B\); thus, we also compute two additional regressions \(I-->N\) and \(I-->B\), in order to compute the Sobel test for the presence of a mediating relationship between \(I\), \(N\), and \(B\). \textit{Fig 4A} visualizes this model graphically, and \textit{Fig 4B} plots sample traces of each process.\\


\textit{Fig 4c} then shows the false-positive rate for each individual regression, and the mediation test, as a function of the  \(\phi\) parameter for two types of null interventions: one generated by an AR(\(\phi = 0\)), which we denote as \textit{static latent dynamics} and one AR(\(\phi = 1\)), which we denote as \textit{autocorrelated latent dynamics}. The false positive rates are then shown for four cases: static and autocorrelated latent dynamics over 100 and 1000 trials. In the static intervention case, it can be seen that while false positives in individual correlations can be seen to rise as a function of \(\phi\), the mediation analysis reduces the false positive rate dramatically. Thus, by the additional knowledge of the latent dynamics between the intervention and measured variables being independently sampled, the spurious relationships found in correlational analysis can be avoided altogether by instead testing the hypothesis of a relationship between \(N\) and \(B\), given the latent dynamics \(I\).\\

However, when the latent dynamics between the intervention and \(N\) and \(B\) are assumed to be time-dependent, here done by simulating the intervention regression time series as a random walk, the false positive rate rises as a function of \(phi\), such that when all three processes are generated independently, a significant relationship is concluded in about 40\% of cases when the experiments were simulated with 100 trials. As such, if an experimentally controlled intervention passes through latent dynamics \(I\) prior to causally effecting \(N\) and \(B\), the expected false positive rate for significant mediation tests is dependent on the latent dynamics, which may or may not be observable and/or falsifiable. Additionally, as can be seen in  \textit{Fig 4C}, when the experiment length is increased to 1000 trials, the number of significant mediation results increases, as was the case when correlating two independent random walks. As such, for single time-series the confounding influence of latent dynamics between independent and dependent variables may be exaggerated, rather than mitigated, by collecting more trials. From this analysis, it can be seen that the inclusion of an experimentally controlled variable can reduce the problem of spurious correlations, but only if this variable (or any variables between the intervention and the causal effect on the measured variables) influences the measured processes through a path in the system that is itself not autocorrelated. When the latent dynamics between intervention and measured variables are assumed to be autocorrelated, the null analysis suggests that the problem of spurious correlation may persist. \\
\begin{figure}[!ht]
\centering
\includegraphics[width=.8\textwidth]{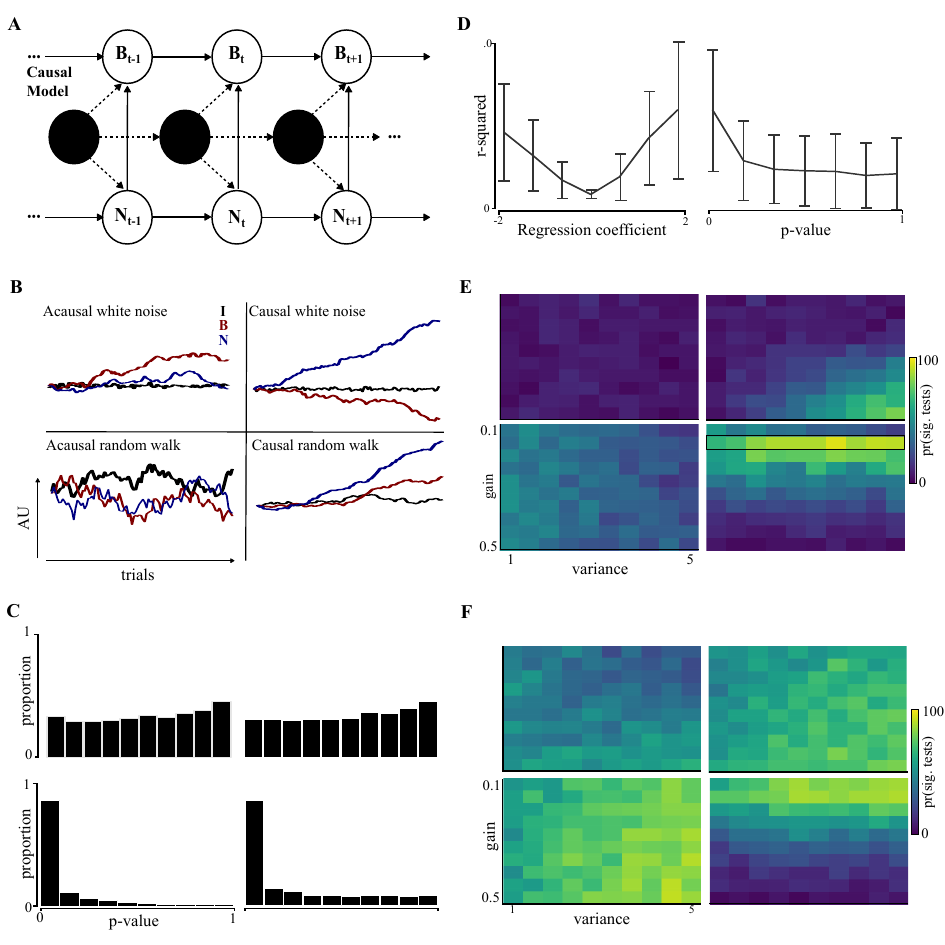}
\captionsetup{justification=raggedright,singlelinecheck=false}
\caption{Latent dynamics control mediation significance testing in a causal effect model. \textbf{A:} Graphical model of the causal effect model. \(B\) and \(N\) are AR-processes auto-correlated according to \(\phi\), and \(N\) effects \(B\) according to a gain factor \(g_{N}\). The latent dynamics of the intervention in turn effect \(B\) and \(N\) by a gain factor \(g_{I}\).\textbf{B}: sample time-series for \(B\), \(N\), and \(I\) for white noise and random walk interventions, where \(g_{I}\) is either 0 (acausal) or non-zero (causal). \textbf{C:} Distribution of mediation p-values for 1000 experiments for causal white noise (top left), acausal white noise (top right), causal random walk (bottom left) and acausal random walk (top right). \textbf{D:} Mean and standard deviation of regression coefficients (left) and p-values are plotted against the intervention to behaviour (\textit{regression c in figure 4)} in 7 bins. \textbf{E:} Proportion of significant tests are plotted as a function of latent intervention dynamics variance (\(sigma_{I}\)) and gain (\(g_{I}\)). Plots are organized as in \textbf{B}. \textbf{F:} Intervention variance-gain plots for the causal random walk model, for different distributions of \(g_{I}\)}
\end{figure}

Defining the latent dynamics of the null intervention allows us to simulate a false positive distribution, but this does not allow us to compare the distribution of mediation p-values to what may be expected in the case where a true effect exists. To address this, we developed a causal mediation model where i) the neural process directly influences the behavioural process at each time-point and ii) the intervention time series directly influences both the behavioural and neural processes. We consider an intervention as a latent dynamic that we specify with an autoregressive process, that in turn influences a neural and behavioural time-series with additive gain \(g_{I}\). To focus on the role of the latent intervention dynamics, we model the neural and behavioural processes as history-dependent, that is a random walk. \(N\) provides additive gain to \(B\), controlled by \(g_{N}\), thus providing a causal effect on the behavioural time series.

\begin{equation}
    \begin{gathered}
    I_t = \phi_1 I_{t-1} + \varepsilon_t \\
    N_t = N_0 + \sum_{i=1}^{t} N_i + I_t *g_I + \varepsilon_t  \\ 
    B_t = B_0 + \sum_{i=1}^{t} B_i + I_t *g_I + N_t * g_N + \varepsilon_t  \\ 
    \end{gathered}
\end{equation}

We illustrate this model graphically in \textit{Fig 5A}, and plot example processes for single experiments in \textit{Fig 5B}, where the latent dynamics of the intervention and gain are changed. We depict four scenarios: a white noise latent dynamic that causally influences \(N\) and \(B\) (that is, \(g_{I} ~= 0\)) denoted \(W_{c}\), a null white noise latent dynamic \(W_{N}\) (\(g_{I} = 0\)) , a random walk intervention that causally influences \(N\) and \(B\), denoted \(R_{C}\), and a random walk null intervention \(R_{N}\) (\(g_{I} = 0\)). We then perform a false positive analysis on the mediation statistic for each model. \textit{Fig 5C} plots the distribution of mediation p-values for each model (top left \(W_{C}\), top right \(W_{N}\), bottom right \(R_{C}\), bottom left \(R_{N}\)). Here, it can be seen that for white noise latent dynamics, a similar distribution of mediation p-values is observed for both the causal and null model, suggesting that assessing causality between \(N\) and \(B\) through \(I\) may lead to similar conclusions regardless of whether or not \(I\) actually influences \(N\) and \(B\). Further, the distribution of p-values is approximately uniform, suggesting that the null hypothesis would likely not be rejected if these experiments could be repeated many times, regardless of whether or not the intervention had a causal effect. \\

For random walk interventions, we see a different trend. Mediation p-values are not uniform, but are heavily skewed towards zero, indicating that over most experiments a significant mediation effect would be reported -- regardless of whether or not an actual effect was observed. To confirm that the reported p-values were representative of the actual regression coefficients in the mediation test, we plotted the mean and standard deviations of the direct regression coefficient \(c\) (intervention onto behaviour) as a function of the variance explained in the regression (\(r^{2}\)) in order to confirm that the coefficients were related to variance explained, and then the mediation p-value compared to the direct regression (\(r^{2}\)), in order to confirm that mediation significance was correlated with variance explained in the direct regression. Overall, this tells us that when using mediation analysis to infer causality between two measured processes \(N\) and \(B\) through the use of an intervention, different latent dynamics in the path \(I\) can obscure true effects when the dynamics are time-independent, ie generated from a white noise process, bu can generate false positives when the latent dynamics of the intervention are history dependent, ie generated from a random walk process. To more fully explore the state space of the model for different interventions, we then systematically varied the intervention gain \(g_{I}\) and intervention variance \(\varepsilon_I\) for each of the four intervention latent dynamics. \\

\textit{Fig 5E} plots a heatmap of mediation p-values for each model variation (top left: acausal-white noise, top right: causal-white noise, bottom left: acausal-random walk, bottom right: causal random walk.)  Here, it can be seen that for the acausal-white noise model, significant mediation results are infrequent, consistent with the latent dynamics of the intervention not influence either \(N\) or \(B\). In this model, the effect of the gain parameter is irrelevant, since no effect is observed, but the variance of the process determines the time-series the regressions are performed on. In the causal white noise model, the proportion of significant mediation tests was dependent on both the variance and the gain of the intervention. Thus, distinguishing between a null and causal intervention was dependent on the statistical properties of the latent intervention dynamics. For random walk latent intervention dynamics (bottom left), the proportion of false positive mediation results for the acausal model increased relative to the acausal white noise model. This suggests that the expectation of different latent dynamics (white noise vs random walk) can determine the expected false positive rate. In the causal random walk model (bottom right), again the gain and variance parameters of the intervention determined the proportion of significant tests. Because of this, for different parameters of the latent intervention dynamics, the causal and acausal model may or may not be able to be distinguished by repeated testing. \\

In the causal random walk model, we observed an interesting nonlinearity in the state-space for very low \(g_{I}\) (<.3), as shown in the black box over \textit{Fig 5E, bottom right}. To explore this further, we simulated the causal random walk model across smaller discretizations of \(g_{I}\). We show the results for this in \textit{Fig 5F}. From here, it can be seen that there is a progressive increase, followed by a decrease, in the proportion of significant mediation tests in this parameter range. Thus, even a simple causal model like this can have interesting nonlinearities in the state-space of latent intervention dynamics, where small changes may elicit different expectations for mediation analysis that would require an \textit{a priori} knowledge of the graphical model and latent dynamics in order to predict. \\

\section*{Discussion}

In this paper, we have argued for the computational undecidability of ground-truth history-dependence, and used model systems where history-dependence can be controlled in order to detail the assumptions about history-dependence required to make causal statements about the relationship between neural and behavioural processes. Given this, a discussion on hypothesis testing in neuroscience is warranted -- rather than provide a barrier to using interventions for understanding the neural basis of behaviour, we argue that explicitly beginning from the starting axiom of irreducible uncertainty would permit a more holistic understanding of the systems under study, and our place in intervening on them. 

\subsection*{Hypothesis testing in absence of ground-truth understanding}
Perhaps unsurprisingly, the problem of spurious regression was first analyzed in the context of econometrics (\cite{granger_spurious_1974}), where trends are not isolated and repeated in a laboratory, but analyzed and forecast as the system, which we concede cannot be controlled experimentally, evolves. A neuroscientist, on the other hand, may object to the characterization of spurious correlation arising from history-dependence as a major concern for their experimental approaches. In doing so, they may explain the distinction between regressing single time series that only occur once, as is typical in economics, and repeatedly running controlled experiments -- such an appeal would rely on the Central Limit Theorem (\textit{Fig 6}). The central limit theorem guarantees that the distribution of sample means of a sufficiently large and independently drawn random sample from any population, regardless of its original probability distribution, will tend to approximate a normal distribution. This approximation becomes increasingly accurate as the sample size grows larger. For the central limit theorem to be valid, samples must be independent of each other. When this condition is met, it allows us to use the properties of the normal distribution for making statistical inferences and hypothesis testing without requiring the data to arise from a process with known and verifiable statistical properties. \\

As such, an experimentalist may be unconcerned with the undecidability of history-dependence, and the resulting uncertainty over the statistical properties of the processes they are sampling within individuals, if the goal is to discern the effect of an experimental intervention at the population level \footnote{In doing so, one is essentially conceding an inability to identify ground-truth processes in favour of statistical testing of group or individual-level changes in the face of external interventions. It may be optimistic to believe that this is the default (or even dominant) understanding when performing statistical testing in neuroscience experiments -- the conclusions as stated often undermine this belief. More likely, the dominance of statistical philosophies that focus on inference from sample to population are rooted in a confluence of historical developments and hybridizations of competing theories -- the entrenchment of these trajectories into accepted paradigms has been suggested as a sort of ritualistic belief system (\cite{gigerenzer_statistical_2018}). Regardless, we will operate under the assumption that this is generally understood so as to better clarify what sorts of hypotheses are justifiable under the appeal to the central limit theorem.}. By collecting enough participants and enough data, they can perform statistical tests on the sample means across different conditions under assumptions of normality, as the distribution of the sampling mean converges to a normal distribution. To be sure, many variables follow a normal distribution at the population level, despite being caused in all likelihood by a complex set of history-dependent factors that may be individually described by probability distributions of arbitrary form. So in appealing to the central limit theorem, an experimenter may argue that, unlike time series that are generally accepted to be non-repeatable, such as economic trends, in the laboratory they can isolate specific situations and repeatedly measure them. Although this requires conceding a general inability to converge on identifying a ground-truth generative process, it allows for sampling to be exploited in order to test hypotheses about experimental interventions at population levels. \\

\begin{figure}[!ht]
\centering
\includegraphics[width = .8\textwidth]{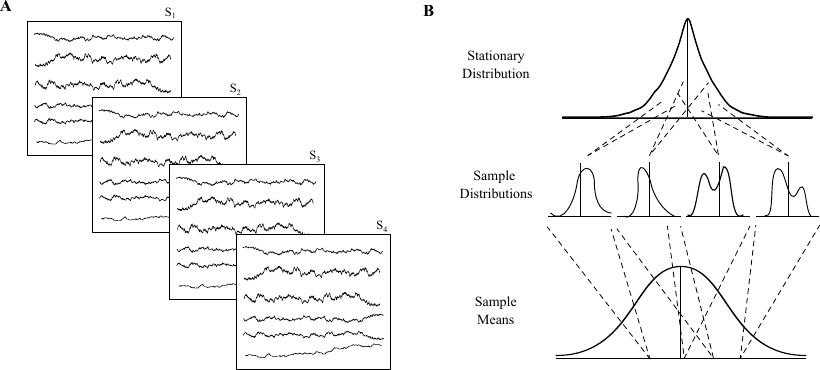}
\captionsetup{justification=raggedright,singlelinecheck=false}
\caption{Central limit theorem allows for test-retesting of effects in non-ergodic processes. A: Data matrix of a hypothetical non-ergodic process, where each row is a single 'trial', and each column represents a time point within the trial. For arguments sake, each data matrix here is collected from an independent process. B: resulting distributions of activity averaged across all trials for each independent matrix. These are cartoons, the point is they are non-Gaussian. With a sufficient number of independent sessions, the distribution over sample means across sessions approaches a normal distribution under the 
central limit theorem.}
\end{figure}

Given the inability to falsify the dynamics of open systems from repeated sampling, it is worth being explicit in what hypotheses are actually testable in absence of a ground-truth understanding of a system. Take for example the testing of how an external intervention changes the group-level relationship between a measured neural and behavioural variable. For this class of question, one would typically design an experiment such that a behaviour could be atomized into independent units, often called trials, where a continuously measured neural variable is aligned to and discretized alongside (\cite{huk_beyond_2018, williams_statistical_2021, stokes_importance_2016}). The goal of statistical analysis for such an experiment is to test whether an intervention changes the correlation structure between neural activity and behaviour at the population-level. For such a claim to hold, spurious correlations arising from history-dependence in independent time series are not necessarily a problem -- while this may introduce substantial across-subject variability, by considering each subject as an unknown data-generating process, a statistical difference in the pooled means of each process should be detectable by repeatedly sampling the population on and off a given intervention. From this, causal statements about population effects of an intervention can be made without having to be concerned about underlying generative models or spurious correlations -- with the important caveat that statements made about this intervention may have little to no predictive power over individual outcomes. \\

However, many other legitimate hypotheses may be made that are more specific than asking if an intervention changes a population-level phenomenon. In neuroscience and psychology, it is increasingly appreciated that inter and intra-individual differences are a dominant source of variability in data that corrupt many assumptions of traditional statistical analyses (\cite{mangalam_point_2021}). Indeed, these sources of variability may not just be noise that an experimenter should seek to eliminate, but a reflection of the specificity of experimental interventions to inter and intra-individual differences that may yield misleading mixtures of effects when inappropriately aggregated. The basic example is that of an intervention that can have different effects depending on some intervening variable. If the intervening variable is not taken into account, averaging across distinct contexts can produce inappropriate null results. Thus, it is important to generate hypotheses about interventions that do not ask whether it has an effect at a population level, but instead in a subset of individuals, a subset of contexts, or both. \\


In refining our hypothesis space to include conditional-dependencies, we can sometimes sacrifice the power to be agnostic about the functional form of the distributions we sample from, as we end up further removed from data that can be taken to be reasonably independent, and into cases where data being combined come quite clearly from dependent sources. At each level of granularity of hypothesis, we therefore require additional "independent" test-retests to be able to abstract over potentially dependent sources of information in order to avoid the issues with spurious regressions, thus permitting the testing of these more granular hypotheses. \\

We illustrate hypotheses of different granularity in \textit{Fig 7}: \textit{general-group}, where the effect of an intervention is tested at a population level; \textit{conditional-group}, where a group-level effect is hypothesized to only be observed in a subset of contexts in a given dataset; \textit{general-individual}, where only certain individuals are hypothesized to respond to the intervention; and \textit{conditional-individual}, where certain individuals only respond to the intervention in certain contexts. In these hypotheses, different assumptions need to be made about independence for conventional statistical testing to be valid. For general-group assessments, as we have suggested, the underlying distribution of each individual, or any other \textit{post-hoc} splitting of a given data-set, need not follow any distribution in particular, so long as it has some theoretical bound. But consider instead a conditional-group assessment, where an experimenter collecting some joint neural and behavioural data and employing the following logic:

\begin{quote}
    \begin{enumerate}
        \item Intervention \(X\) only influences neural measure \(Y\) under conditions where neural modulator \(T\) is high
        \item Neural modulator \(T\) is only observed to be elevated when a subset of the behaviour \(B_{s}\) is observed
        \item Intervention \(X\) should only influence correlation between \(Y\) and \(B_{s}\), and not \(Y\) and \(B_{\sim s}\)
    \end{enumerate}
\end{quote}

\begin{figure}[!ht]
\centering
\includegraphics{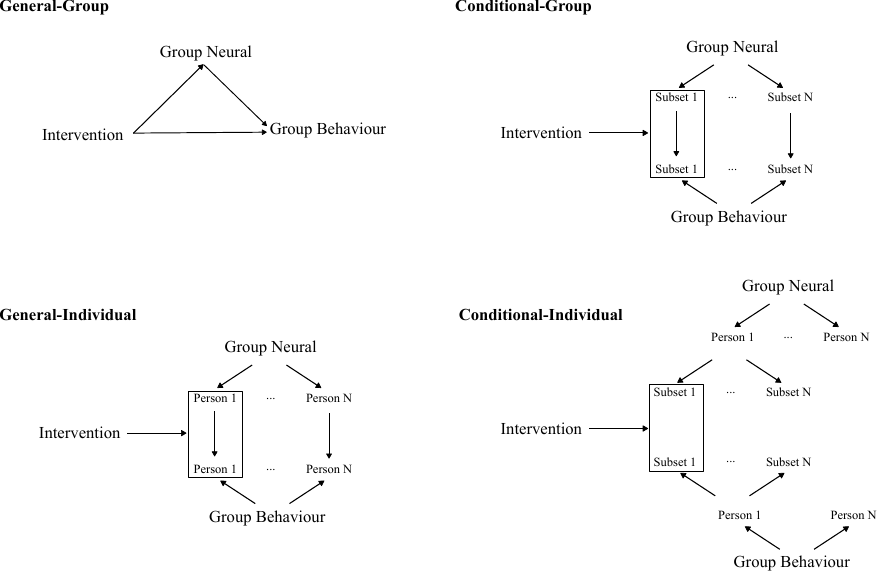}
\captionsetup{justification=raggedright,singlelinecheck=false}
\caption{Granularity of hypotheses on brain-behaviour interventions. A: General-group hypothesis, where a causal intervention is hypothesized to influence a behavioural process through a neural process on average, measured across a population. B: Conditional-group hypothesis, where the same causal intervention is hypothesized to only influence a subset of data in the joint neural-behavioural space. C: General individual, where a causal intervention is hypothesized to only influence a subset of people in a population. D: Conditional individual, where a causal intervention is hypothesized to influence specific people under specific subsets of a larger dataset.}
\end{figure}

This may be a perfectly valid hypothesis to generate. However, in performing hypothesis testing on a subset of the data, we rely on that subset having a stationary distribution that can be sampled from independently of the other  subsets. In short, it introduces the hypothesis of a data subset being a discrete and independent category, without necessarily verifying that statistical inference of differences under this assumption can be justified. To be sure, these sorts of hypotheses may represent reasonable claims, perhaps justified by some prior knowledge about the system. Returning to the example of random walks, imagine studying two independent processes where it is hypothesized that successive repetitions of the same outcome result from a conditionally-independent state than successive alternations. This may be a perfectly valid hypothesis to hold --  as such, one may deconstruct a time series into categories of different sequences, building a sort of library of hypothetical states with different associations to the generative statistics of a task. But how would one go about verifying the ontological validity of such a categorization? In an alternative case of the generative models being described by a purely random process, very similar subsets could likely be demarcated post-hoc, yet in this case we know that they come from a process where i) these classes are not a natural kind, and ii) data in these classes are not conditionally-independent.  \\

Individual variability represents a different level of granularity than conditional-group hypotheses. Many causal interventions to brain-behaviour relationships are thought to be specific to individual physiology and circumstance, but hypotheses in neuroscience and psychology are typically tested by collecting large participant pools (\cite{molenaar_new_2009, fisher_lack_2018}). In the case where a hypothesis about an intervention \textit{is} specific to an individual, a new granularity of independence must be proposed in order to avoid the spurious correlation problem of regressing single time series. Here, one may again rely on the atomization of the trial, such that many independent measures can be collected within a single session. In this case, one could formulate hypotheses for specific individuals, or sets of individuals that follow some pattern in a population, once again relying on the assumption that these trials are in fact independent. \\

Given a survey of the literature on sequential dependencies in behaviour, a researcher may well deem this to be an unacceptable assumption, and conclude that a history-dependence in the behaviour may introduce the problem of spurious correlation we observed in analyzing independent random walks. Instead, they may conclude that in order to study person-specific (sometimes called low-N designs) brain-behaviour relationships, many different repeated measures of \textit{sessions} are required (as B.F. Skinner said, “It is more useful to study one animal for 1000 hours than to study 1000 animals for one hour (\cite{graham_small_2012})). By performing such a study, they would hope to abstract over dependencies between trials by getting an estimate of the session-session variability. However, such a study would pose an additional hypothesis: are there dependencies between sessions? It is easy to see here how recursive logic could be employed to iteratively question the claims for independence, propose an experiment that requires additional data at a timescale further removed, then question its independence or propose a correction \textit{ad infinitum}. The result is a never-ending chain of plausible hypotheses for the functional role of a behaviour spanning the summed length of an experiment, requiring an experiment of an additional length to treat statistically, and so on. \\

Practically, at some point we have to choose whether data are functionally independent enough for statistical testing. Although truly demonstrating independence may be computationally undecidable, we may be content in saying that some things are more dependent than others. For example, ignoring the statistical challenges of sequential dependencies across trials in an experiment with a learning effect, or averaging the effects of a neural intervention across individuals with different relevant physiology, may be deemed to be unreasonable. On the other hand, the dependencies of seasonal variables on test-retesting within an individual over many months may not be so important in determining the effect of an intervention that acts over minutes. However, in between such extremes lies much room for interpretation. What principles should guide our decision-making when assessing the relative independence of brain-behaviour data in response to interventions? If we are forced to concede that explicit falsification of ground-truth internal dynamics is undecidable, we have shown here that true effects can be masked, and false effects exaggerated, depending on our knowledge of the underlying graphical model. Uncertainty about the context of an intervention with respect to the ongoing internal dynamics of the system under study thus poses a challenge to the notion that collecting more data leads to more certainty about the effect of an independently controlled variable, particularly if the system under study is non-ergodic. \\

In this paper, we have shown that this uncertainty is a fundamental property of open-systems that cannot be isolated from exchanging information with the world, and analyzed the statistical properties of stochastic processes with different causal structures in order to demonstrate the ease with which different latent dynamics may obscure real effects or amplify the relationship between independent signals. As a result, we conclude that causal inference in neuroscience should necessarily begin with uncertainty over possible causal models, which although are almost certainty unobservable, may be more or less likely at different levels of analysis.

\section*{Code Availability}
Code to run the simulations is available at \url{https://github.com/BJCaie/Ergodicity/}

\section*{Acknowledgements}
This research was undertaken thanks in part to funding from the Natural Sciences and Engineering Research Council of Canada (NSERC), the Canada Foundation for Innovation (CFI), and the Connected Minds Program, supported by Canada First Research Excellence Fund, Grant CFREF-2022-00010.

\printbibliography

\end{document}